# Single-mode Dispersion-engineered Nonlinear Integrated Waveguides for Ultra-broadband Optical Amplification and Wavelength Conversion

Ping Zhao[1,2✉], Vijay Shekhawat[1], Marcello Girardi[1], Zonglong He[1], Victor Torres-Company[1], and Peter A. Andrekson[1✉]

[1]Photonics Laboratory, Department of Microtechnology and Nanoscience, Chalmers University of Technology, Gothenburg 41296, Sweden.

[2]College of Electronics and Information Engineering, Sichuan University, Chengdu 610065, China.

✉zhao.ping@scu.edu.cn
✉peter.andrekson@chalmers.se

**Abstract**

Four-wave mixing is a nonlinear optical phenomenon that can be utilized for wideband ultra-low-noise optical amplification as well as for wavelength conversion. This has extensively been investigated for various applications such as communications, spectroscopy, metrology, quantum computing and bio-imaging. However, there is a clear desire to implement these functionalities in a small footprint nonlinear platform, being compatible with large volume fabrication and being capable of efficient operation across a large optical bandwidth. Many such platforms, e.g., silicon, aluminum gallium arsenide, silicon nitride, nonlinear glass, etc., have been explored, but suffer from intrinsic significant performance degradation, i.e., gain and bandwidth deduction and distortion, because conventional approaches of nonlinear photonic waveguide geometry construction for dispersion engineering focus on waveguide cross section and result in always being multimode as a byproduct. Here we propose and demonstrate a methodology that utilizes not only the impact of the waveguide cross section on the modal and dispersion behavior of the waveguide but also includes the impact of the waveguide bend for cutting off high-order modes. This approach results in simultaneous single-mode operation and dispersion engineering for very broadband operation of four-wave mixing. While we implemented this in silicon nitride waveguides, which has emerged as a promising platform capable of continuous-wave optical parametric amplification, the design approach can be universally used with other platforms as well. By carefully also considering both second- and fourth-order dispersion we achieve unprecedented amplification bandwidths of approximately 300 nm in silicon nitride nonlinear waveguides of which the losses can be as record-low as 0.6 dB/m. In addition, penalty-free all-optical wavelength conversion of 100 Gbit/s data in a single optical carrier over 200 nm is realized, for the first time, without optical amplification of signal or idler waves. These single-mode hyper-dispersion-engineered nonlinear integrated waveguides can become practical building blocks in versatile nonlinear photonic devices and optical networks.

**Main text**

**Introduction.** With the distinct advantages of overcoming the bandwidth, noise figure and wavelength range of a stimulated-emission based optical amplifier and generating waves beyond conventional lasers, four-wave mixing (FWM) has led to numerous applications in the fields of communication [1][2], computing [3][4], spectroscopy [5], metrology [6], imaging [7][8], aerospace [9] and quantum optics [10][11]. Particularly, hyper dispersion engineering, i.e., second- and fourth-order dispersion in tandem [12], is quite critical for broadband FWM with parametric gain which is pursued in various areas, such as ultra-long-haul signal transmission [13][14], all-optical high-speed signal processing [15], light detection and ranging (LiDAR) [16], widely-tunable coherent wave generation [17], biochemistry analysis [18] and continuous-variable quantum computing [11]. Since the invention of low-loss silica fibres which provides long nonlinear optical interaction distances, FWM in waveguides has been intensively investigated [19]. Although silica optical fibres are widely available, they exhibit low nonlinearity and narrow parametric bandwidth, suffering from polarization and dispersion drifts which further cause decreases in both FWM efficiency and bandwidth [12][20]. Featured with high nonlinearity, small footprint and flexible patterning, $\chi^{(3)}$-based nonlinear integrated waveguides exhibit excellent on-chip control of optical field properties, such as nonlinear coefficient, dispersion and polarization state, potentially paving way to high-efficiency and wideband FWM [21][22]. Semiconductor nonlinear integrated waveguides with small refractive index contrast attracted much interest for FWM in early times, while they were subject to limited dispersion engineering and severe interferences from other nonlinear effects [23]. Advances in deposition, crystal cutting and bonding have enabled the creation of high-refractive-index-contrast nonlinear integrated waveguide structures with silica cladding, featuring strong field confinement and dispersion engineering [24], ideally suited for broadband parametric signal processing. Vast kinds of such nonlinear platforms bloomed for FWM [25], including silicon [26]–[29], silicon nitride [30]–[32], aluminum gallium arsenide [17][33][34][35], nonlinear glasses [36], [37], graphene [38][39], and plasmonic [40]. Recently, continuous-wave (CW) optical parametric gain was achieved for the first time in $Si_3N_4$ nonlinear integrated waveguides with the balance in propagation loss, nonlinearity, power handling ability and dispersion engineering [30], which also corresponds to high conversion efficiencies (CEs, the power ratio of output idler to input signal).

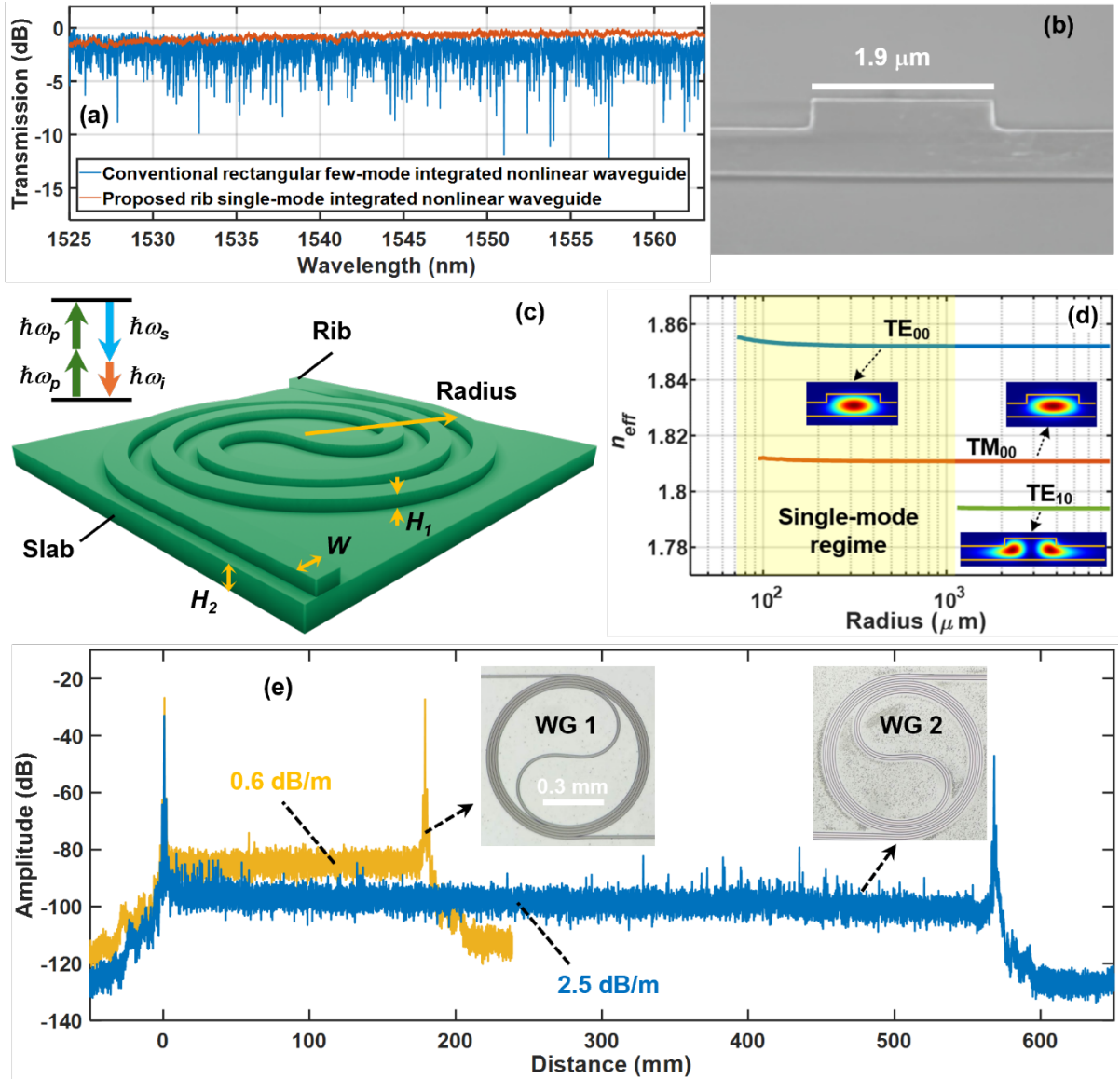

**Figure 1. Principle of single-mode dispersion-engineered nonlinear integrated waveguides for ultra-broadband optical amplification and wavelength conversion.** (a) Normalized measured transmission spectra of a conventional rectangular-core (blue) and a proposed single-mode rib (red) $Si_3N_4$ nonlinear integrated waveguides (WGs). The fabrication of the few-mode rectangular-core spiral $Si_3N_4$ nonlinear integrated waveguide was optimized to reduce sidewall roughness. The wavelength tuning step in the measurements is 1 pm. Both waveguides are about 50 cm long. (b) Scanning electron microscope image of the cross section of one proposed single-mode rib $Si_3N_4$ nonlinear integrated waveguide. (c) Schematic diagram of the proposed rib nonlinear integrated waveguide with proper bend radius for low-loss single-mode operation. $SiO_2$ is the cladding. (d) Effective refractive index of different modes in a rib $Si_3N_4$ nonlinear integrated waveguide varying with the bend radius, with $W$ = 1.9 μm, $H_1$ = 300 nm and $H_2$ = 500 nm. The blue, red and green lines are for $TE_{00}$, $TM_{00}$ and $TE_{10}$ modes, respectively. The insets are the intensity profiles of different modes, and the yellow lines are the $SiO_2$-$Si_3N_4$ boundaries. (e) OFDR traces of 18-cm (yellow, WG 1) and 56-cm (blue, WG 2) long single-mode rib $Si_3N_4$ nonlinear integrated waveguides with wavelength scanning from 1480 nm to 1640 nm. The minimal propagation loss of WG 1 and WG 2 are 0.6 dB/m and 2.5 dB/m. The two insets are the optical microscope images of one spiral unit of WG 1 and WG 2 on two wafers, respectively.

Regarding pump-degenerated FWM, a strong pump (*p*) and a week signal (*s*) waves enter a $\chi^{(3)}$-based nonlinear optical medium where the signal is amplified and an idler wave is generated at an angular frequency of $\omega_i = 2\omega_p - \omega_s$, as illustrated by Fig. 1(a). The phase mismatching parameter $\Delta K = \sum_{k=1}^{\infty} 2\beta_{2k} \Delta\omega^{2k}/(2k)! + 2\gamma P$ affects the gain, CE and bandwidth, where $\beta_i$ is the *i*-th-order derivative with respect to the angular frequency $\omega$ of the optical propagation constant $\beta$ evaluated at the pump frequency, $\gamma$ is the nonlinear coefficient, $\Delta\omega$ is the angular frequency difference between the pump and signal waves and *P* is the pump power [12]. Anomalous dispersion ($\beta_2 < 0$) is usually used to realize phase matching ($\Delta K = 0$) [2]. Conventional high-index-contrast silica-cladding nonlinear integrated waveguides only address transverse cross section to achieve anomalous dispersion at the pump wavelength for parametric gain and high CE [17], [30], [31], [37], while they are always multimode as a byproduct. However, random unavoidable modal coupling results in power drops in signal and pump waves, causing not only a decrease in FWM gain, CE and bandwidth but also distortion of modulated signals [41]. For instance, the blue curve in Fig. 1(a) is the measured normalized transmission spectrum of a conventional rectangular-core dispersion-engineered $Si_3N_4$ integrated nonlinear waveguide of which the fabrication was optimized to reduce the sidewall roughness. The waveguide is 2000 nm wide, 690 nm high and about 50cm long, supporting 7 modes in total. As can be seen in Fig. 1(a), this typical conventional $Si_3N_4$ integrated nonlinear waveguide suffers from serious and dense spectral fluctuations (power fading at some wavelengths more than 10 dB) due to random modal coupling. Single-mode $\chi^{(3)}$-based high-index-contrast silica-cladding nonlinear integrated waveguides with simultaneous anomalous dispersion are strongly desired for efficient wideband FWM, but has never been reported so far, which has been a long-term challenge [24]. Moreover, hyper-dispersion engineering, i.e., $\beta_4$ in tandem with $\beta_2$, is quite important to ultimately broaden the FWM bandwidth. Nevertheless, ultra-broadband single-mode high-CE FWM assisted hyper-dispersion engineering is still missing in integrated nonlinear platforms to date.

**Investigation and results.** Here, we propose a waveguide design method of combining the longitudinal bending with the transverse cross section construction, i.e., three-dimensional geometry, to address the above issues. The red curve in Fig. 1(a) is the measured transmission spectrum of one waveguide we proposed and fabricated (**Methods**) which exhibits excellent single-mode property in contrast to the conventional rectangular-core integrated nonlinear waveguide. Rib waveguides with silica cladding are used to achieve less guiding modes as well as lower propagation losses, compared to rectangular-core waveguides with the same width and total thickness [42]. Figure 1 (b) shows the shows the scanning electron microscope image of the cross section of a proposed 1.9-μm-wide single-mode Si$_3$N$_4$ rib nonlinear integrated waveguide. The key technique to achieve single-mode operation is bending the waveguide to cut off higher-order modes. Figure 1 (c) shows the schematic diagram of the proposed waveguide. $W$, $R$, $H_1$ and $H_2$ are the rib width, radius, height, and slab thickness, respectively. To verify the effectiveness of the proposed method, we use the Si$_3$N$_4$ integrated platform as an example to realize single-mode dispersion-engineered nonlinear waveguides. Figure 1(d) presents the simulated effective refractive index ($n_{eff}$) of different modes as a function of waveguide radius at 1550 nm wavelength. In the simulation, the rib Si$_3$N$_4$ nonlinear waveguide was 1.9-μm wide with $H_1$ = 300 nm and $H_2$ = 500 nm. The blue and red lines correspond to the fundamental transverse-electric (TE) and transverse-magnetic (TM) modes. As can be seen in Fig. 1(d), this rib Si$_3$N$_4$ nonlinear waveguide supports three modes when it is straight. When the radius decreases to less than 1150 μm, the high-order mode TE$_{10}$ (the green line) is cut off. Hence, we obtain a single-mode-per-polarization spiral rib Si$_3$N$_4$ nonlinear waveguide with proper bend arrangement in principle. With a larger nonlinear coefficient than TM$_{00}$ mode, TE$_{00}$ mode is considered for FWM. The TE$_{00}$ mode dispersion at 1550 nm can be anomalous despite waveguide bend and tuned by changing the rib width (Fig. E1(a)). Moreover, the proposed single-mode rib Si$_3$N$_4$ nonlinear integrated waveguide exhibits good fabrication tolerance on second- and fourth-order dispersion considering typical thickness variations (+/- 3 nm) and width uncertainties (+/- 10 nm) in fabrication (Fig. E1(c) and (d)). Moreover, figure 1(e) shows the traces of optical frequency-domain reflectometry (OFDR) of two TE$_{00}$-mode-coupled 1.9-μm-wide spiral rib Si$_3$N$_4$ nonlinear integrated waveguides with lengths of 18 cm (yellow, WG 1) and 56 cm (blue, WG 2). Although the slab widths of WG 1 and WG 2 are different, they are sufficiently large so that $TE_{00}$ mode is not affected by the slab sidewall (**Supplementary**). The measured propagation loss of WG 1 is 0.6 dB/m, which is the lowest among all the high-confinement passive integrated photonic waveguides. For WG 2, the propagation loss is about 2.5 dB/m and varies along the waveguide length. The local small optical scattering of WG 2 is more frequent than WG 1. The insets in Fig. 1(e) are optical microscope images of WG 1 and 2, respectively, where one can see that WG 2 suffers from residual nanoparticles of which WG 1 is almost free. The propagation loss difference between WG 1 and 2 is mainly owing to the fabrication variations which we are trying to improve. We fabricated six 56-cm-long single-mode rib $Si_3N_4$ nonlinear integrated waveguides and WG 2 is the only one without big defects on the OFDR traces. The yield of 18-cm-long rib waveguides is 4/20, mainly limited by the misalignment of the dual-layer tapers and minor defects.

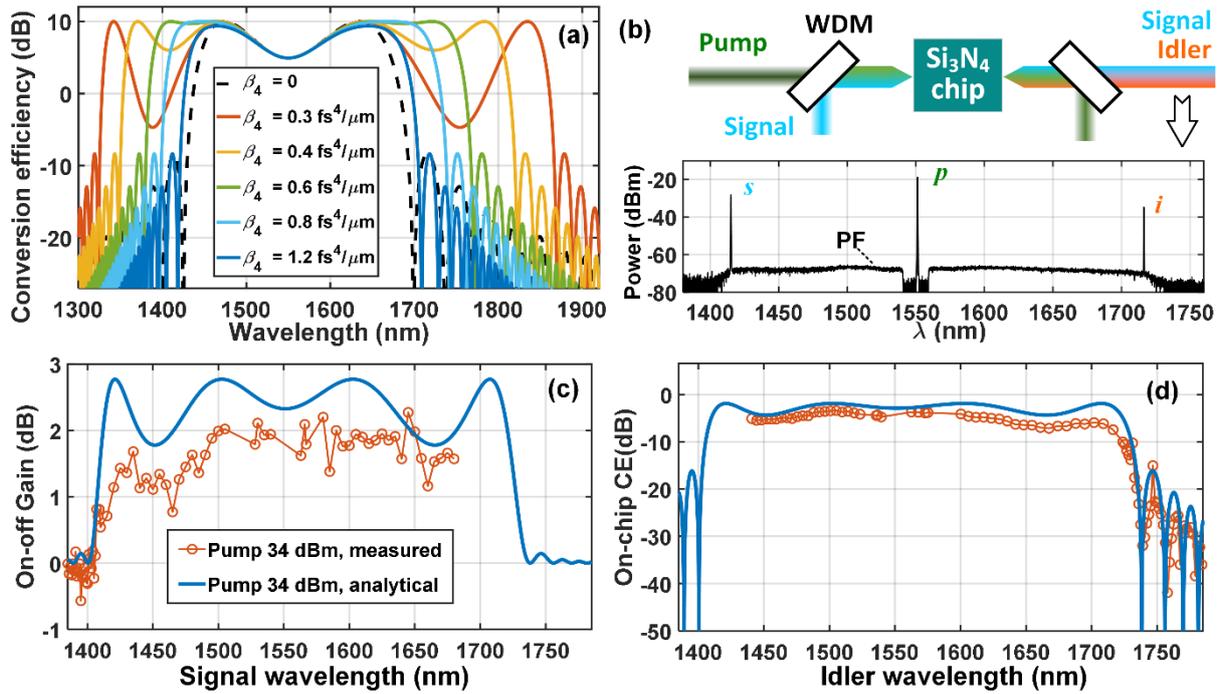

**Figure 2. Ultra-broadband integrated parametric waveguides with hyper-dispersion engineering.** (a) Theoretical conversion efficiency spectral curves of 1-m-long $\chi^{(3)}$-base nonlinear integrated waveguides with various fourth-order dispersions, respectively. The waveguide loss is 1 dB/m with a nonlinear coefficient of 0.7 (Wm)$^{-1}$ and second-order dispersion of -1 ps$^2$/km. The power at 1550 nm wavelength is 35 dBm. The red, yellow, green and blue solid lines are for the cases of $\beta_4$ = 0.3 fs$^4$/μm, 0.4 fs$^4$/μm, 0.6 fs$^4$/μm and 1.2 fs$^4$/μm, respectively. The black-dash line corresponds to the case where fourth-order dispersion is not considered. (b) Experimental diagram for the FWM characterization of the 0.56-m-long single-mode rib Si$_3$N$_4$ nonlinear integrated waveguide. The inset is a spectrum after the WDM coupler with a 1551.1 nm pump and a 1415 nm signal. WDM, wavelength division multiplexing; PF, parametric fluorescence. (c) Measured (red circle) on-off parametric gain and (d) on-chip CE of the single-mode Si$_3$N$_4$ nonlinear integrated waveguide a 34-dBm on-chip pump power. The solid blue curves are calculated analytically with the assumption of a spectrally constant waveguide loss of 2.5 dB/m.

Apart from the single-mode waveguide property, we investigated how the hyper-dispersion engineering ultimately extends the bandwidth of FWM-based parametric process. Fig. 2(a) depicts the theoretical CE spectra of a 1-m-long nonlinear integrated waveguide pumped by a CW 35-dBm wave at 1550 nm. In the calculation, the 2$^{nd}$-order dispersion was –1 ps$^2$/km at 1550 nm, the effective nonlinear coefficient was 0.7 (Wm)$^{-1}$ and the waveguide loss was 1 dB/m. The maximum CE reaches 10 dB and corresponds to a maximum parametric gain of about 10 dB for signal wave [43]. The parametric gain spectrum is similar to the CE spectral curve in this case. The black-dash line is for $\beta_4 = 0$ in which case the amplification bandwidth is 270 nm. The red, yellow, green and blue solid lines are for the cases of $\beta_4 = 0.3$ fs$^4$/µm, 0.4 fs$^4$/µm, 0.6 fs$^4$/µm and 1.2 fs$^4$/µm, respectively. As shown by Fig. 1(e), the amplification bandwidth increases to 542 nm for $\beta_4 = 0.3$ fs$^4$/µm since the fourth-order dispersion leads to new phase-matching wavelengths [12]. When the balance among the nonlinear shift, second- and fourth-order dispersion is achieved with $\beta_4 = 0.6$ fs$^4$/µm, two flat gain regimes are obtained and the amplification bandwidth reaches 385 nm, i.e., 43% bandwidth increase compared to the case of $\beta_4 = 0$. Hence, fourth-order dispersion plays a vital role in realizing ultra-wideband parametric devices. Furthermore, we characterized the ultra-broadband FWM in WG 2 using CW pump-probe approaches based on the experimental diagram shown by Fig. 2(b) (**Method**). The on-chip pump power was 34 dBm, considering the coupling loss. The inset in Fig. 2(b) is an output optical spectrum of WG 2 with the residual pump mitigated by a wavelength division multiplexing (WDM) coupler, where the signal, pump and idler wavelengths are 1415 nm, 1551.1 nm and 1716 nm, respectively. Ultra-wideband flat parametric fluorescence (PF) during FWM is also observed, as can been seen in Fig. 2(b), which may find applications in metrology [44][45] and quantum optics [10]. Figure 2 (c) and (d) depict the measured (blue) on-off parametric gain and on-chip CE spectra, respectively. On-off gain is used here since it can mitigate the impact of wavelength-dependent coupling loss of the tapers on the measurements. The solid lines are theoretically fitted spectra with $\beta_2 = -2.2$ ps$^2$/km and $\beta_4 = 1.9$ fs$^4$/µm at 1551 nm. The measured and theoretical curves are in good agreement with small discrepancies which may be due to the wavelength-dependent loss of the waveguide. The on-chip waveguide loss in the L band is about 1 dB, indicating that we achieve 1-dB on-chip net CW parametric gain. Besides, we obtain a maximum on-chip efficient CE of -3.4 dB at 1500 nm wavelength, as shown in Fig. 3 (d). According to the gain and CE spectra in Fig. 2 (c) and (d), we realize a FWM bandwidth of 330 nm, i.e., the widest among all reported CW optical amplifiers to date. Since there were not enough lasers to cover the full FWM bandwidth during the measurements, we recorded the pure PF spectrum indicating the parametric gain profile and checked how the PF spectral shape changes with the dispersion by adjusting the pump wavelength (**Supplementary**). Moreover, the fitted second- and fourth-order dispersion agrees with the waveguide design, and this verifies that the $Si_3N_4$ nonlinear integrated platform is well tolerant to fabrication uncertainties.

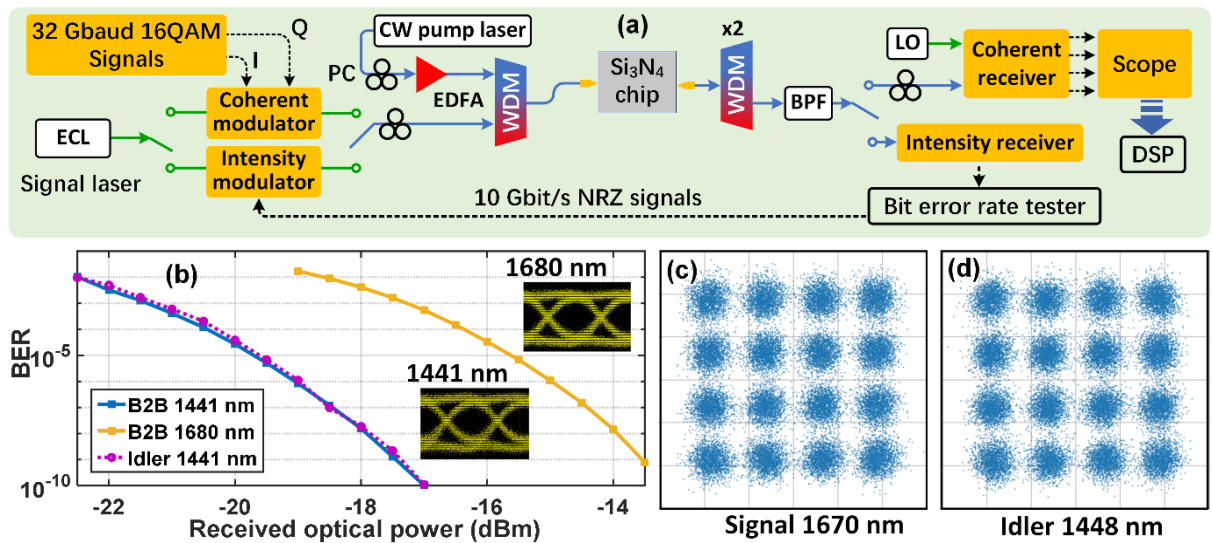

Figure 3. Ultra-broadband high-efficiency high-speed all-optical wavelength conversion based on single-mode spiral rib $Si_3N_4$ nonlinear integrated waveguides. (a) Experimental setup of the $Si_3N_4$-chip-based all-optical wavelength conversion for both intensity-modulation and coherent optical communications. The on-chip pump regarding the coupling losses was 34 dBm. ECL, external-cavity laser; BPF, band-pass filter; LO, local oscillator. (b) Bit error rate as a function of received optical power of 10-Gbit/s non-return-to-zero (NRZ) signals (1680 nm) and converted idlers (1441 nm, purple-dashed). The insets show the eye diagrams of the back-to-back (B2B) optical signals at wavelengths of 1441 nm and 1680 nm, respectively. (c) 1670-nm signal and (d) 1448-nm idler constellation diagrams with 32-GBaud 16-QAM modulation after offline digital signal processing (DSP).

**Application.** Furthermore, we applied the $Si_3N_4$-chip-based ultra-broadband efficient FWM to all-optical high-speed wavelength conversion (WC) for communications. Figure 3(a) shows the experimental setup where intensity and coherent modulation are both included (**Methods**). We used 10-Gbit/s non-return-to-zero (NRZ) intensity modulation to check the WC impairments to the idler quality. Figure 3(b) presents the bit-error rate (BER) of back-to-back (B2B) 1441 nm (blue solid), 1680 nm (yellow solid) signals and 1441 nm idler (purple dotted), respectively. The B2B BER difference between 1441 nm and 1680 nm signals are due to the wavelength-dependent responsivity of the intensity receiver. As can be seen in Fig. 3(b), the idler suffers negligible penalty compared to the 1441 nm B2B signal, which indicates the proposed CMOS-compatible single-mode $Si_3N_4$ nonlinear integrated waveguide is promising in all-optical signal processing. In addition, we implemented the all-optical wavelength conversion of single-polarization 32-GBaud 16-quadrature-amplitude-modulation (16-QAM) signals with a net rate over 100 Gbit/s, based on the 56-cm-long single-mode $Si_3N_4$ integrated nonlinear waveguide. Figure 3 (c) and (d) show the constellation diagrams of B2B 1670 nm signal and converted 1448 nm idler, respectively. This is the first time that more-than-200-nm-wide single-wavelength 100-Gbit/s-beyond all-optical WC without the amplification of signal/idler wave is realized. As the 32-GBaud 16-QAM is the dominant modulation format of current optical fiber communication systems connecting the continents on Earth, the $Si_3N_4$-chip-based high-efficiency WC demonstrated has a bright future in all-optically reconfiguring global WDM optical networks, unlocking C&L-beyond transmission bands of optical fibres [46][47] and increasing the capacity and flexibility of optical neuromorphic computing for artificial intelligence [4].

**Discussion and conclusion.** Ultra-broadband optical amplification is a basic and appealing area in photonics. Figure 4 summarizes the bandwidth of various types of wideband CW optical amplification. Stimulated emission has been widely used for optical amplification over decades.

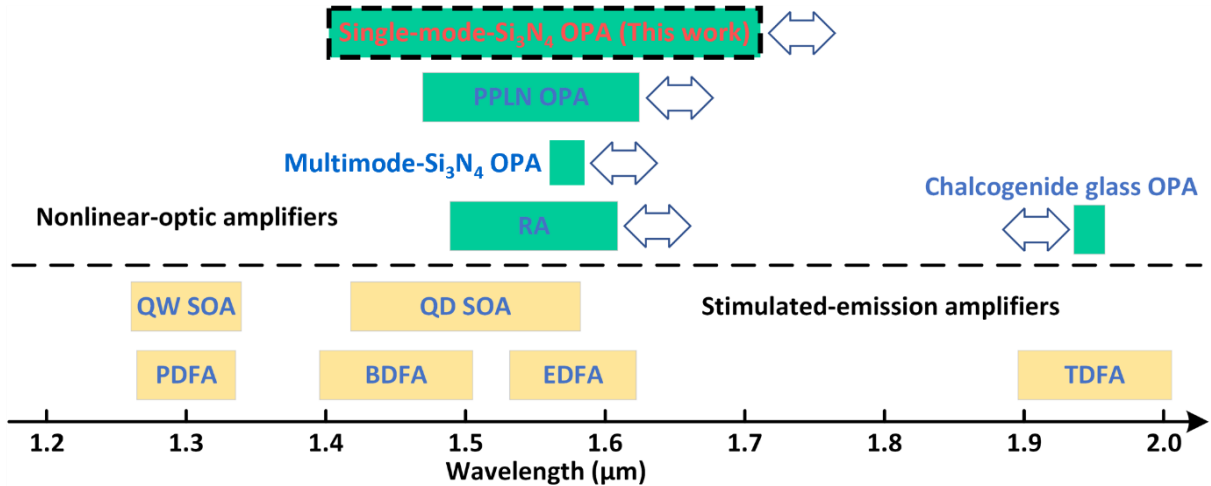

**Figure 4.** State-of-the-art bandwidth of CW optical amplification in the near-infrared regime based on both stimulated-emission and nonlinear-optic platforms.

Various material platforms were developed to construct stimulated-emission optical amplifiers, such as EDFA [48], bismuth-doped fiber amplifier (BDFA) [49], thulium-doped fiber amplifier for optical (TDFA) [50], praseodymium-doped fiber amplifier (PDFA) [51], quantum-well (QW) and quantum-dot (QD) semiconductor optical amplifiers (SOAs) [52][53]. To cover the whole transmission window of telecommunication silica and widely-concerned hollow-core fibres [54], new stimulated-emission materials with different bandgaps need to be investigated and developed, and this is a long-term task and consumes vast resources. On the other hand, nonlinear optical effects including Raman [55], difference frequency (DF) and FWM provide insights to build wideband optical amplifiers of which the operation wavelength can be flexibly tuned by pump frequency and waveguide design without changing host materials. Amplifiers based on the DF or FWM effects are also generally called optical parametric amplifiers (OPAs). Periodically poled lithium niobate (PPLN) waveguide OPAs based on DF effect were widely investigated [56], leading to ultra-high-speed optical fiber transmission beyond conventional telecommunication bands [57]. However, PPLN nonlinear waveguides are subject to significant challenges, such as difficulties in fabrication (especially for low-loss artificial periodic poling), dispersion engineering for ultra-wide band and high temperature sensitivity. Based on the FWM process, OPAs using $\chi^{(3)}$-nonlinear nanophotonic waveguides are free of periodic poling and have the advantages of ease in fabrication and dispersion engineering. CW nanophotonic OPAs were first realized with chalcogenide glasses microfibers [58], while they may suffer from power handling issues for high gain [59]. Benefitting the hyper dispersion engineering in single-mode rib $Si_3N_4$ nonlinear integrated waveguides in this work, an OPA bandwidth of 330 nm is demonstrated, for the first time, and much wider than those of the other CW optical amplifiers demonstrated to date, as shown in Fig. 4(a). Currently, the length of the fabricated 1.9-μm-wide single-mode spiral rib $Si_3N_4$ integrated nonlinear waveguide greatly limits the gain as well as the FWM CE. We are continuing the fabrication optimization and believe that the yield of long ultra-low-loss single-mode rib $Si_3N_4$ nonlinear waveguides will be improved in future. The length issue together with the yield may not be problematic for CMOS foundries where deep-ultra-violet exposure technologies are mature for massive nanofabrication.

Considering a propagation loss of 0.6 dB/m with a pump power of around 34 dBm, a maximum gain of about 20 dB can be expected if the waveguide length can be 2 meters. With the advances in semiconductor optoelectronics, Watts CW pump lasers can also be available [60] and may lead to compact OPAs via hybrid photonic integration [61]. The OPA spectral flatness over the 200 nm bandwidth can be improved via dual-pump, dispersion or pump-phase shifting techniques which can be implemented in integrated waveguide platforms [62][63].

In this paper, we proposed and demonstrated record-low-loss single-mode $Si_3N_4$ nonlinear integrated waveguides with hyper dispersion engineering for ultra-broadband efficient CW FWM. Distinct from conventional approaches focusing on waveguide cross-section, we exploited the three-dimensional waveguide geometry parameters for on-chip optical field manipulation, simultaneously achieving single-mode transmission and dispersion engineering of nonlinear integrated nanophotonic waveguides in telecommunication bands for the first time. Using the 0.56-meter-long single-mode rib $Si_3N_4$ nonlinear integrated waveguide with hyper dispersion engineering, we obtained a record CW gain bandwidth of 330 nm in the near infrared regime. The whole transmission window of telecommunication silica fibres can be potentially covered by the parametric gain profiles of single-mode rib $Si_3N_4$ nonlinear integrated waveguides integrated on the same wafer. Additionally, we realized the widest all-optical WC of single-wavelength 100-Gbit/s-beyond signals without amplifying the signal/idler wave. These experimental results agree with theoretical expectations. Higher CW parametric gain and CE can be expected with longer low-loss single-mode rib $Si_3N_4$ nonlinear integrated waveguides. With the combination of cross-section shaping and longitudinal bending, the waveguide technique we proposed is easy to implement and universal for other integrated platforms to realize low-loss single-mode dispersion-engineered nonlinear waveguides that can be key blocks of photonic devices such as lasers[64][65], switches [66], resonators [67] and promising from fundamental research of photonics, physics, quantum, chemistry and biology to industrial applications in communications, computing, spectroscopy, imaging and metrology.

**MATERIALS AND METHODS**

**Fabrication and linear characterization of spiral rib waveguides**

The method to manufacture the proposed rib waveguide was based on subtractive E-beam lithography fabrication process of ultralow-loss high-confinement $Si_3N_4$ waveguides [68][69]. An 800-nm-thick $Si_3N_4$ layer was deposited with low pressure chemical vapor deposition (LPCVD) on a 4-inch Si wafer with 3-μm-thick $SiO_2$ layer on top. Two-step etching was used to fabricate the rib waveguides with dual-layer tapers at the chip edges for the coupling with lensed fibres. The 300-nm-thick 1.9-μm-wide $Si_3N_4$ spiral rib was defined in the first etching. Three-micrometer-wide micro grooves between the rib waveguides were formed in the 500-nm-thick slab layer during the second etching to prevent the cracking of $Si_3N_4$. The etched waveguides were then annealed above 1100 °C in Ar-flow atmosphere and cladded with 3-μm-thick $SiO_2$ via LPCVD. Finally, the whole wafer was diced into chips via etching. The parameters of WG 1 and 2 are listed in Table 1.

Table 1. Parameters of single-mode spiral rib $Si_3N_4$ nonlinear integrated waveguides

| Waveguide | Length (cm) | Rib gap (μm) | Slab width (μm) | Minimum bending radius (μm) | Maximum bending radius (μm) | Chip size (mm) |
|---|---|---|---|---|---|---|
| WG 1 | 18 | 15 | 12 | 180 | 450 | 1 x 14 |
| WG 2 | 56 | 50 | 30 | 165 | 450 | 3 x 29 |

Lensed fibers with beam-spot diameters of 3 um were used to couple light with the $Si_3N_4$ nanophotonic chip. We found that the average coupling loss was about 2.5 dB/facet at 1550 nm wavelength for the $TE_{00}$ mode of the spiral rib $Si_3N_4$ waveguide. The waveguide propagation loss was measured using a commercial OFDR tester combined with a wavelength-scanning laser. Regarding the linear transmission spectrum measurements of the $Si_3N_4$ waveguides, we focused on the $TE_{00}$ mode and used a power-constant tunable laser with a wavelength step of 1 pm.

**FWM characterization with single-mode rib $Si_3N_4$ nonlinear integrated waveguides**

We used a pump-probe approach of measuring the parametric gain and CE of the CW FWM in the 0.56-meter-long single-mode rib $Si_3N_4$ nonlinear integrated waveguide. Three semiconductor external-cavity lasers (ECLs) were utilized to generate a signal wave that can be tuned from 1355 nm to 1680 nm. Another semiconductor ECL emitted a 1551.1 nm pump wave which was amplified by a high-power EDFA. The pump and signal waves were combined by a low-loss thin-film WDM coupler with a bandwidth of 4 nm and entered into the single-mode rib $Si_3N_4$ nonlinear integrated waveguide via the lensed fibre. The polarization states of both the pump and signal waves were aligned to the $TE_{00}$ mode of the rib $Si_3N_4$ waveguide. One percent of the optical field at the $Si_3N_4$ waveguide input port was recorded by an optical spectrum analyzer (OSA). At the $Si_3N_4$ waveguide output port, we used a 15-nm-wide coarse WDM coupler to mitigate the residual pump intensity before we measured the optical spectrum. With power calibrations for the input and output optical spectra, we calculated the FWM gain and CE for each signal wavelength.

**Wavelength conversion for optical communications**

Based on the 0.56-meter-long single-mode rib $Si_3N_4$ nonlinear integrated waveguide, we implemented all-optical wavelength conversion of NRZ and 16-QAM signals. The pump wavelength was also 1551.1 nm with a CW on-chip power of 34 dBm in all the measurements. No optical amplification was used in the signal/idler wave paths. The green and blue lines in Fig. 3 (a) are polarization-maintaining (PM) and non-PM single-mode fiber patch cords, respectively.

For the NRZ modulation format, a Mach-Zehnder modulator (MZM) was utilized to convert the 10-Gbit/s electrical signals from a bit error rate tester (BERT) into optical domain with a carrier wavelength of 1680 nm. The optical power of the MZM output signal was about 2 dBm. After a 10-Gbit/s NRZ idler wave at 1441 nm was generated during the on-chip FWM process, two coarse 1550 nm WDM couplers with bandwidths of 15 nm were used to well mitigate the residual pump. We used a band-pass filter (BPF) to select the idler wave. A 10-GHz intensity receiver with one photodetector and two radio-frequency amplifiers converted the optical signals back to electrical domain and fed to the BERT to calculate the BER and record the eye diagrams of the received signal.

Regarding the 16-QAM optical signals, an electrical arbitrary waveform generator was used to generate 32-GBaud in-phase and quadrature components which were amplified separately and sent to a single-polarization coherent electrical-optical modulator. To detect the 16-QAM optical signals after wavelength conversion, a commercial coherent receiver with another tunable ECL as a local oscillator was used. The signal wavelength at the chip input was set to be 1670 nm so that the 16-QAM data could be converted to a 1447 nm idler wave at which the wavelength-dependent coherent receiver responsivity can still be sufficient for data recovery. The electrical signals after the coherent receiver were recorded by a high-speed real-time scope. Off-line digital signal processing was applied to analyzing the signal BER and constellation.

**Acknowledgements**

The authors would like to thank Prof. Magnus Karlsson and Yan Gao for fruitful technical discussions, Estrella Torres, Mats Myremark and Dr Chao Lei for their help on chip preparation and experiments, Prof. Minhao Pu at Technical University of Denmark, Frida Olofsson, and Santec Corporation for sharing equipment. This work was performed in part at Myfab Chalmers.

**Grants**

This work was partially funded by the Swedish Research Council (grant VR-2015-00535 to P.A.A. and grant VR-2020-00453 to V.T.-C.), the K.A. Wallenberg Foundation and KAW Scholar (P.A.A.).

**Author contributions:**

Conceptualization: P.Z.

Methodology: P.Z. designed the waveguide and performed the chip characterization. V.S. and M.G. developed the waveguide nanofabrication. P.Z., Z.H. and V. S. conducted chip applications of optical transmission.

Investigation: P.Z. performed the theoretical modeling of waveguides and the simulation of the nonlinear process, designed the waveguide layout. V.S. fabricated the waveguides. P.Z. performed linear and nonlinear characterization of the waveguides. P.Z., Z.H. and V.S. implemented the wavelength conversion experiment for optical communications.

Data processing: P.Z.

Funding acquisition: P.A.A. and V.T.-C.

Project administration: P.A.A. and V.T.-C.

Supervision: P.A.A.

Writing (original draft): P.Z.

Writing (review and editing): P.A.A., V.S., and V.T.-C.

**Data availability**

The data that support the findings of this study are available from the corresponding author upon request.